\documentstyle[12pt,aasms4]{article}

\journalid{337}{}
\articleid{11}{14}

\def\spose#1{\hbox to 0pt{#1\hss}}
\def\lta{\mathrel{\spose{\lower 3pt\hbox{$\mathchar"218$}}
     \raise 2.0pt\hbox{$\mathchar"13C$}}}
\def\gta{\mathrel{\spose{\lower 3pt\hbox{$\mathchar"218$}}
     \raise 2.0pt\hbox{$\mathchar"13E$}}}
\def\n{\noindent}

\def\be{\begin{equation}}
\def\ee{\end{equation}}
\def\msun{M_{\odot}}

\def\mdot{\dot M}

\def\mpy{M_{\odot} yr^{-1}}

\begin{document}

\title{Rapid Bursts From GRS 1915+105 with RXTE}
\author{Ronald E. Taam\altaffilmark{1}, Xingming Chen\altaffilmark{2}, 
and Jean H. Swank\altaffilmark{3} }

\n \altaffilmark{1}{Department of Physics \& Astronomy, Northwestern 
University, Evanston, IL 60208
\vskip -0.12in
taam@ossenu.astro.nwu.edu}

\n \altaffilmark{2}{UCO/Lick Observatory, Board of Studies in Astronomy 
and Astrophysics, University of California, Santa Cruz, CA 95064
\vskip -0.12 in
chen@ucolick.org}

\n \altaffilmark{3}{NASA, Goddard Space Flight Center, Greenbelt, MD 20771
\vskip -0.12in
swank@pcasun1.gsfc.nasa.gov}

\begin{abstract}
The light curves of GRS 1915+105 observed with RXTE on October 15, 1996 reveal 
a wide range of transient activity including regular bursts with a recurrence 
time of about 1 minute, irregular bursts, and dips. In contrast to bursts from
other sources, a secondary (and a tertiary) weaker burst immediately 
following the primary burst are observed. Detailed energy spectra indicate 
that the source softens during the main outburst and successively hardens 
during the secondary and tertiary bursts. This may imply that the accretion 
flow has a corona-disk configuration and that the relative contribution of the 
hot corona decreases during the primary bursts and increases during the 
secondary and tertiary bursts. The primary burst profile resembles the bursts 
produced in the time dependent evolutions of accretion disks which are 
thermally and viscously unstable. The secondary burst may reflect an inward 
shift of the inner edge of the disk which results in a greater release of 
gravitational binding energy.  
\end{abstract}


\section{INTRODUCTION}

The hard X-ray transient source, GRS 1915+105, was discovered by Castro-Tirado, 
Brandt, \& Lund (1992) in 1992 using the WATCH all sky monitor instrument 
onboard the GRANAT satellite.  It is located at a distance of about $12.5 \pm 
1.5$ kpc (Mirabel \& Rodriguez 1994) with an outburst X-ray luminosity of $\sim 
10^{39}$ ergs s$^{-1}$.  In contrast to previous transient sources, GRS 1915+105
is distinguished by the fact that it ejected radio components with apparent 
superluminal motion up to $v/c = 1.25$ (Mirabel \& Rodriguez 1994).  

GRS 1915+105 was also observed to be X-ray active in 1994 using the BATSE 
instrument on CGRO (Harmon et al. 1994; Greiner et al. 1994) and using the 
SIGMA telescope on GRANAT (Finoguenov et al. 1994). It has been
found to be in outburst in 1996 based on observations obtained by 
the Rossi X-ray Timing Explorer (RXTE). The RXTE observations of GRS~1915+105 
have revealed a wide range of transient activity such as the occurrence of 
dips, irregular bursts (Greiner, Morgan, \& Remillard 1996; Belloni et al.
1997; Swank, Chen \& 
Taam 1997), and quasi-periodic oscillations (Chen, Swank, \& Taam 1997; 
Morgan, Remillard, \& Greiner 1997).  On October 15, 1996, the 
source entered into an unusual phase of activity in which X-ray bursts were 
emitted in a quasi regular pattern.

In this Letter we report on the properties of these quasi regular bursts. 
In \S 2 the light curves, temporal structure, energy spectra in the burst 
active and burst inactive state, and spectral fits to these spectra are 
presented. The interpretation of the bursts in terms of thermal instabilities 
in the accretion disk and the implications of the spectral fitting are 
discussed in the final section.

\section{OBSERVATIONS}

On October 15, 1996 GRS 1915+105 was observed for a total time of about 7.2 hr 
with the Proportional Counter Array (PCA) onboard RXTE. In these observations, 
the mean count rate of the source in the 2 - 13 keV energy band was 
$\sim 15000$ counts s$^{-1}$.  On this day, the source was situated on the soft 
branch of the hardness-intensity diagram as described in Chen et al. (1997).

\subsection{Burst Profiles}
  
During a period of about 5.5 hours the source emitted a series of quasi regular 
bursts.  The light curve (in the 2-13 keV band) for a portion of this period 
is illustrated in Figure 1. An important feature is the secondary burst which 
always follows the primary burst. The occurrence of bursts (both primary and 
secondary) is quasi regular with recurrence timescales ranging from 60 to 
100~s. The transition to the primary burst occurs gradually on the shoulder 
of the light curve preceding the onset of the main burst. The count rate at 
peak to that on the shoulder is about a factor of 2 higher, exceeding
$\sim 20000$ counts s$^{-1}$.  Note that the post burst count rate is 
significantly reduced to about 6000 counts s$^{-1}$.  

The details of the burst profile in the 2 - 13 keV band are shown in Figure 2a 
with a higher time resolution. 
The burst exhibits variable oscillations on shorter
timescales, which makes the estimate of the burst timescales 
(rise, decay, duration) difficult. Therefore the variations of the burst 
timescales with respect to recurrence timescale are not easily quantified.
The primary burst is characterized by more 
structure with variability on a timescale of one second. The main burst has a 
slow rise ($\sim$ 8 s), a fast decline ($\sim 2$ s), and a duration of 
$\sim 10$ s. The secondary burst occurs about 3 s later, it is weaker (with 
a peak count rate $\sim$ 35\% lower than the main burst), has a comparable 
duration, and is characterized by a quasi symmetric rise and decay time of 
$\sim 2.5$ s. The time between the peaks of the primary and the secondary 
bursts is about 10 s. In some cases, there is evidence for a much weaker 
tertiary burst.  The peak count rate of the tertiary burst is lower than 
the secondary burst by about 35\% - 50\%. 

To examine the variations in burst profile with energy, we plot the 
corresponding light curve in the 13 - 60 keV band in Figure 2b. It is seen 
that the main burst in the 2-13 keV band is now very weak or does not exist 
at all. On the other hand, the secondary burst and, especially, the  
tertiary burst in the 2-13 keV band is very pronounced in the hard energy band.
We also note the existence of oscillations (with time scale
about 3-10 seconds) which have an amplitude of about $20\%$, much larger
than the corresponding oscillations in the 2-13 keV band. 
However, the power density spectra of these time series show only broad 
features and do not reveal any narrow QPO peaks.
  
The hardness ratio defined as the count rate in the 13-60 keV to 2-13 keV 
band is illustrated in Figure 2c.  It can be seen that the hardness ratio 
decreases during the main outburst and recovers during the post outburst 
phase. It reaches a peak ($\sim 0.09$) at the luminosity peak of the tertiary
burst. Notice that, the $\sim 3-10$ s oscillations are seen in 
the ratio plot, suggesting a different oscillation strength in these
two energy bands. 

\subsection{Spectra and Energies}

Energy spectra have been calculated using recently determined 
detector response matrices (Version 2.1.2) and background software 
(version 1.4g of the "pcabackest" tool and versions 1.1, 3.0, and 1.0 of 
the cosmic ray, the activation, and the X-ray backgrounds respectively). 
Data from the modes B\_2ms\_4B\_0\_35\_H were used for 4 channels below 13.1 
keV and from the mode E\_16us\_16B\_36\_1s for 16 channels above 13.1 keV at 
the gain of the epoch of the observation. To investigate the spectral 
evolution during the bursts, as guided by the ratio plot (computed with 62.5 
ms resolution) spectra were determined with 0.625~s resolution on the shoulder
prior to the primary burst, at the peaks of the primary burst, the secondary 
burst, and the tertiary burst. For the shoulder, it was also feasible to 
use the Standard-2 mode data with 16 s time resolution and 129 energy 
channels. The energy spectra for the four stages are 
illustrated in Figure 3a.  The spectra are qualitatively similar, 
but cross at about 15 keV. The ratios of the spectra are sensitive to the 
differences, as is seen in Figure 3b,
where the ratios are calculated with 
respect to the shoulder spectrum. The spectra are softer during the 
peaks of the main and second outbursts in comparison to the shoulder of the 
light curve, whereas the spectrum during the third peak is the hardest.  

The best fits to a number of single component spectral models (power law, 
bremsstrahlung, exponentially cut-off power law, various representations 
of Comptonization) were much poorer than the combination of a
multi-temperature disk black body plus a power law (diskbb + 
powerlaw in XSPEC, Mitsuda et al. 1984). A $2\%$ systematic error on 
the model was assumed in the fitting with the most current response 
matrices because the fits to the Crab spectra imply such errors in the
response in the range 3-7 keV and greater than that above 30 keV. 
With the systematic error the reduced $\chi^2$
for the fits to the spectra summed from 6 bursts 
with single component models was 4 (versus 70 without the systematic error), 
while it was 1.5 for the disk black body and power law combination. 
The binned and event mode data gave results for the shoulder consistent with 
those for the higher energy resolution Standard-2 data, but with the 
parameters of the disk component more accurately determined, as expected.

The results are summarized in Table 1. The inferred color
temperatures in the inner region of the disk increase from 1.3 keV at the 
shoulder to 2.4 keV at the peak of the secondary burst, but decrease to 
$\sim 2.2$ keV at the tertiary burst peak. The photon 
indices are comparable ($\sim 3.3$) at the primary and secondary peaks, but  
lower ($\sim 2.9$) at the tertiary burst peak. The inferred luminosities 
of both components vary. In the tertiary peak, a small (inner radius) 
disk is indicated, but at a relatively low confidence. The 
parameters are different by $10-20\%$ for different versions of 
the response matrix, but the relative nature of the fits does not change. 
In comparison to our $R_{in}$ ranging from $32 \pm 2$~km to $8 \pm 3$~km, 
Belloni et al. (1997) obtain a minimum $R_{in}$ of $20.3 \pm 0.3$~km for 
data on Oct. 7, 1996 and a maximum of $319 \pm 9$~km. 

Replacing the multi-temperature disk black body with a simple one temperature 
black body component (bbodyrad in XSPEC) gives nearly as good a fit.
The parameters of the power law 
component are nearly independent of this choice. The black body temperature is
about $30\%$ lower than the hottest temperature for the disk, and
the inferred radius is about twice that of the disk model. While in 
principle, effects of scattering and or radiation pressure could be 
included in models, available models are not consistent or agreed upon. 
The multicolor black body model is simple, gives good fits to 
many black hole candidate spectra, and provides a standard of 
comparison (See discussion by Ebisawa 1994). 

\section{DISCUSSION}

GRS 1915+105 was found to be in a new bursting state during which it emitted 
quasi regular bursts with a period ranging from 60 - 100 s.  The sharp decay 
and the more gradual recovery of these bursts resemble the burst profiles seen 
in numerical calculations of thermal/viscous instabilities in accretion 
disks as reported by Taam \& Lin (1984) (see also, Lasota \& Pelat 1991; 
Cannizzo 1996).  Within this framework the recovery phase corresponds to the 
diffusion of matter into the region which has been depleted by the previous 
outburst.  An example of such a calculation is illustrated in Figure 4. Based
upon a model consisting of a cool disk and corona and in which the viscosity 
in the disk is proportional to the total pressure (see Chen 1995; Abramowicz, 
Chen \& Taam 1995), bursts are produced which resemble those observed, 
particularly, the broad shoulder and narrow peak features are 
reproduced.  For an accretion rate of $4 \times 10^{-8} \mpy$ and 
a black hole mass of $10 \msun$, the best fit disk parameters are an 
$\alpha $ viscosity parameter of $\sim 0.1$ and a dissipation of 85\% of the 
gravitational binding energy in a corona.  The bursts in the 
theoretical light curve are characterized by a recurrence time of about 100 s 
and a ratio of peak to shoulder and peak to post dip count rates of 2.5 and 5 
respectively. The region participating in the outburst involves radii less 
than about $3 \times 10^8$ cm from the central object.
We note that, in Belloni et al. (1997) a much smaller viscosity
parameter, $\alpha \sim 0.01$, is required to fit the dip time scale.

Theoretical models that we have calculated over a 
range of a factor of 10 in accretion rate about $4 \times 10^{-8} \mpy$, 
in which the corona is absent do not reproduce the observed light 
curve because the unstable region is too wide leading to larger amplitude 
outbursts and the lack of a significant shoulder.  Hence, the 
thermal instabilities based upon radiation pressure effects in an optically 
thick disk (Lightman \& Eardley 1974; Shakura \& Sunyaev 1976) without a 
corona are insufficient for detailed fitting to the observational data in a 
quantitative manner for these accretion rates. Similar burst
timescales can be produced for lower mass objects with a corresponding 
reduction in $\alpha$.  However, the luminosity level is then too
low to be consistent with that observed ($> 10^{39}$ erg s$^{-1}$ 
at a distance of 12.5 kpc).

The detailed spectra of the bursts suggests that the accretion flow can also be 
interpreted within a picture of a hot corona and a cool disk component. That 
is, for a fixed fraction of gravitational energy dissipated in the corona 
(as suggested by the fit to the burst shape), the
spectral evolution during the burst reveals that the relative contribution 
from the cool disk increases during the main outburst and the contribution due 
to the hot corona increases during the post main outburst state. In addition,
as is seen from Table 1, the power law index is smaller in the shoulder state 
while the peak black body temperature is higher in the burst state.

A feature of the bursts from GRS 1915+105, which the models do not
capture, is the presence of a secondary burst emitted about 10 s after the
main burst.  We considered whether this burst could be explained as a 
consequence of radiation
feedback on the corona-disk configuration due to the main outburst. In
particular, if the radiation emitted from the inner regions  cooled the
corona in the external region immediately neighboring the inner regions
involved in the main outburst, it would  destabilize the underlying cool 
disk (see Ionson \& Kuperus 1984).  
The radial extent of the unstable region would  increase as mass
in these regions diffuses to the compact object. However, the spectral
fitting (see Table 1) does not support this model. Instead, it suggests an
inward shift of the inner edge of the disk, which then results in a
greater release of gravitational binding energy. 

An inward shift of the inner edge of the disk is possible, which would lead 
to a greater release of gravitational binding energy, provided that   
the accretion flow was  advection dominated before reaching the horizon
of the black hole. In other words, the radial extent of the advection dominated
flow region is decreased.  Chen \& Taam (1996) showed that, as the
mass accretion rate decreases, the inner edge of the disk shifts inwards.
In fact, the third burst can be explained similarly with the difference that 
the dissipation in the corona increases, which
results in a harder spectrum and a smaller power law index. 
The shift of the inner disk edge would occur at the local diffusion 
timescale of the innermost region of the disk, which is less than
a second for a reasonable viscosity parameter ($\alpha \sim 0.1$).

The occurrence of the quasi regular pattern is not a sole function of the 
source intensity since during other observations where the mean count 
rate was also about 15000 counts s$^{-1}$, the behavior was different, sometimes
without significant oscillations (e.g. August 18) and sometimes with larger
oscillations but less regular (e.g. June 16).  
It is possible that in those cases the source did not remain at the required 
mass accretion rate level for a sufficiently long time to enter into the rapid 
burst state. 

The smallest radius for the inner edge (ie., the location of the sonic 
point) of the disk lies between the marginally bound and the last 
stable orbits of a test particle, which is 
$6 M/\msun < R_{in}^{\rm min} ({\rm km}) < 9 M/\msun $
for a non-rotating Schwarzschild, and $R_{in}^{\rm min} ({\rm km})
\gta 1.5 M/\msun$ for a maximumally rotating Kerr black hole of mass $M$.
If one interpretates the 67 Hz QPO in terms of the 
disk oscillation models (Chen \& Taam 1995; Morgan \& Remillard 1996; Milsom 
\& Taam 1996; Morgan, Remillard, \& Greiner 1997; Nowak, Wagoner, Begelman, 
\& Lehr 1997), then the black hole mass implied is either $10\msun$
(Schwarzschild) or $33\msun$ (Kerr), which thus gives a minimum radius
of $\sim 50-60$ km independent on the rotation of the black hole.
The disk radii obtained in the spectral fits are underestimates since 
effects due to scattering 
in the disk atmosphere would imply an effective temperature which may be 
smaller by a factor of 1.7 (Shimura \& Takahara 1995). This "hardening factor" 
would lead to an increase in the radius by a factor of 3, that is, 
$R_{in} \sim 24-39$ km. In addition, if the effective emission area of the 
disk is less than the surface area of the disk, in the case where the disk is 
patchy or has ring-like structures, the radius can be even larger (e.g., 
Haardt, Maraschi, \& Ghisellini 1994). 
Thus, we conclude that the inner radius deduced from the spectral fitting
is likely to be a lower limit, and is consistent with GRS 1915+105 being 
either a non-rotating ($\sim 10\msun$) or a rotating ($\sim 30\msun$) 
black hole.

\acknowledgements

\n This research was supported by NASA under grants NAGW-2526 and
NAG5-3059, by NSF grant AST-9315578, and by the 
RXTE NRA-1 grant 10258 through the University Space Research Association
(USRA) visiting Scientist Program.

\newpage
\begin{figure}
\caption{Light curve of GRS 1915+105 on 15 October 1996 revealing
periodic outburst behavior.  The data corresponds to the 2-13 keV count
rate in the 5 XTE/PCA detectors.  Note that the main outburst is followed by
a smaller secondary outburst.}
\end{figure}

\begin{figure}
\caption{(a) Upper panel, expanded view of one of the bursts in Figure 1.
The horizontal scale of the light curve is enlarged to
reveal the detailed burst profile.  (b) Middle panel, the corresponding
light curve in the energy band of 13 - 32 keV.
(c) Lower panel, the hardness ratio 
corresponding to the ratio of the count rate at 13 - 32 keV to that at 
2 - 13 keV.}
\end{figure}

\begin{figure}
\caption{ (a) Left panel, energy spectra during the four stages of a burst.
The square, triangle, asterik, and circle are for the spectra on the primary
peak, the second peak, the tertiary peak, and the shoulder respectively.
(b) Right panel, spectrum ratio with respect to the shoulder.
The square, triangle, and asterik are for the the primary
peak, the second peak and the tertiary peak respectively.}
\end{figure}

\begin{figure}
\caption{  The theoretical light curve of the disk corona system based upon 
a model disk extending to $10^9$ cm from a black hole. 
Note the presence of a shoulder before the burst and the post burst
dip.}
\end{figure}

\clearpage
\begin{table*}
\begin{center}
\begin{tabular}{cccccccc}
\tableline
Location& $T_{in}$(keV) & $R_{in}$(km) & $\Gamma$ & K &$L_{disk}$
&$\mdot_{disk}$ & $L_{PL}$\\
Shoulder  & $1.35\pm 0.04$ & $30\pm 2$ & $3.02\pm 0.03$
& $60\pm 5$ & $0.40 $ & $0.93 $& $1.7E^{-1.04}$\\
Peak 1  & $1.73 \pm 0.03$ & $32\pm 2$ & $3.30\pm 0.08$
& $93\pm 23$& $1.22 $ & $3.1 $ & $2.1E^{-1.30}$\\
Peak 2  & $2.44 \pm 0.04$ & $13\pm 1$ & $3.24\pm 0.07$
& $63\pm 13$& $0.81$ & $0.84$  & $1.5 E^{-1.24}$\\
Peak 3  & $2.20 \pm 0.15$ & $8\pm 2$ & $2.88\pm 0.06$
& $34.5\pm 6$ & $0.19 $ & $ 0.12 $ & $1.2E^{-0.88}$
\end{tabular}
\end{center}
\caption{Best fit parameters: The column absorption $\rm N_H$ in
$10^{22}\, \rm cm^{-2}$ (assumed to be the same at the four stages of the
burst), was $6.3\pm 1.0$ with version 2.1.2 of the response 
matrices. $T_{in}$ and $R_{in}$ are obtained
from the disk-blackbody spectrum, they represent respectively the
temperature
and the radius at the inner edge of the disk. $\Gamma$ (dimensionless)
and K (in photons/keV/cm$^2$/s) are the photon index and the coefficient
of the power law. The bolometric disk-blackbody luminosity ,
$L_{disk}=4 \pi R_{in}^2 \sigma T_{in}^4$, and
the power law component luminosity of energy above $E$, $L_{PL}$, are in the
unit of
${10^{39}\,\rm ergs\, s^{-1}}$, and $E$ is in unit of keV.
We have assumed that the distance to the source is $d=12.5$ kpc
and the inclination angle is $i=70$ degrees (Mirabel \& Rodrigues 1994).
$\mdot = 8 \pi R_{in}^{3} \sigma T_{in}^{4} / 3GM$ is given in units of 
$10^{-8} \mpy$ for the optically thick disk alone, for a compact object 
mass of $10 \msun$. 
}
\end{table*}

\end{document}